\documentclass[reprint,amsmath,amssymb,aps,prd,twocolumn]{revtex4}

\usepackage{graphicx,color}
\usepackage{dcolumn}
\usepackage{bm}
\usepackage{soul}

\begin{document}
 
\preprint{APS/123-QED}

\title{Solar gamma ray probe of local cosmic ray electrons}

\author{Hong-Gang Yang$^{1,2}$}
\author{Yu Gao$^{3}$}
\author{Yin-Zhe Ma$^{4,5,1}$}\thanks{Corresponding author: Y.-Z. Ma, mayinzhe@sun.ac.za}
\author{Roland M. Crocker$^{6}$}
\affiliation{
$^1$Key Laboratory of Radio Astronomy, Purple Mountain Observatory, Chinese Academy of Sciences, 
Nanjing 210023, China\\
$^2$School of Astronomy and Space Science, University of Science and Technology of China, Hefei, Anhui 230026, China \\
$^3$Key Laboratory of Particle Astrophysics, Institute of High Energy Physics, Chinese Academy of Sciences, Beijing, 100049, China\\
$^4$Department of Physics, Stellenbosch University, Matieland 7602, South Africa \\
$^5$National Institute for Theoretical and Computational Sciences (NITheCS), Stellenbosch, Matieland, 7602, South Africa\\
$^6$Research School of Astronomy and Astrophysics, Australian National University, Canberra ACT 2611, Australia}

\date{\today}

\begin{abstract}
TeV-range cosmic ray electrons and positrons (CREs) have been directly measured in the search for new physics or unknown astrophysical sources. CREs can inverse-Compton scatter solar photons and boost their energies into gamma ray bands. Any potential CRE excess would enhance the resultant inverse Compton emission spectrum in the relevant energy range, offering a new window to verify the measured
CRE spectrum. In this paper, we show that an excess in the TeV range of the CRE spectrum, such as the one indicated by the DAMPE experiment, can induce a characteristic solar gamma ray signal. 
Accounting for
contamination from extragalactic gamma ray backgrounds (EGB), we forecast the DAMPE feature is testable ($\gtrsim 4 \sigma$) with a $\sim 10^{5}\,\mathrm{m}^2\,{\rm yr}$ exposure in the off-disk direction.
This can be achieved by long-exposure observations of water Cherenkov telescopes, such as LHAASO (7.2 years) and HAWC (25.9 years).
\end{abstract}

\maketitle


\section{Introduction}


While propagating through the Milky Way, TeV electrons can lose energy quickly via radiative cooling mediated by synchrotron emission and inverse-Compton scattering (ICS) with interstellar radiation fields (ISRFs). Therefore, local measurements of TeV cosmic ray electrons and positrons (CREs) are sensitive probes of the presence and transportation of electrons in the Galaxy. In particular, fast cooling means that the approximation of the continuous source distribution for electrons can break down if a nearby cosmic ray source exists. The recent CRE measurement from the DAMPE experiment revealed an excess signal at $\sim 1.4\,{\rm TeV}$~\cite{DAMPE_Nature} with an estimated $2.3\sigma$ global significance and locally at more than $3\sigma$~\cite{Fowlie:2017fya}. The origin of this excess is unclear, and a number of possibilities are discussed in Ref.~\cite{Q_Yuan2017}, including undiscovered new sources (see e.g. Refs.~\cite{Q_Yuan2017,Gao2020} for theoretical interpretations). Notably, the measured CRE spectrum can vary between different datasets, such as from DAMPE, AMS02~\cite{AMS:2014bun}, FermiLAT~\cite{Fermi-LAT:2017bpc} and CALET~\cite{Adriani:2018ktz}. Therefore, an independent measurement will be of great interest to offer a complementary test on any TeV CRE spectral feature, including the 1.4 TeV excess.

High-energy CREs can kick solar photons up to the gamma ray band through ICS, generating a halo of gamma ray emission around the Sun (denoted as the halo component in the following).
~\citet{OS2007} and ~\citet{IC_Moskal} showed that the halo component cannot be neglected if measuring diffuse Galactic gamma ray emission (DGE) and the extragalactic gamma ray background (EGB). 
In fact, the initial 
evidence for the halo component was found in archival data of EGRET~\cite{EGRET_data}.
The halo was clearly resolvable 
from the pointlike gamma ray emission of the solar disk induced by cosmic ray cascades in the solar atmosphere using 1.5-year Fermi-LAT data~\cite{Abdo_2011}. 
The spectrum of the halo component covers a wide energy range from MeV up to the TeV band ~\cite{StellarICS}. 
ICS photons partially inherit spectral features of the incident CR electrons. The halo component's intensity is also expected to vary due to the modulation effect on the CRE flux induced by the solar wind and magnetic field. Therefore, measuring the spectrum of the halo component can shed some light on the CRE spectrum and solar modulation in the entire heliosphere~\cite{Abdo_2011,StellarICS}. 

In this paper, we propose using the halo ICS component as a cross-test of the spectrum of TeV-range CREs. As an example, we use the TeV excess in the CRE spectrum suggested by DAMPE to calculate the off-disk solar gamma rays spectrum, and forecast the detectability of the excess signal given the backgrounds. We will calculate the required exposure time for water Cherenkov telescopes such as such as HAWC~\cite{HAWC_1} and LHAASO~\cite{ALHAASO} to 
achieve
such a detection. Our formalism is not only applicable to the particular DAMPE TeV excess behavior, but to the general case of a CRE excess signal.


\section{Methods}
When considering the photon field close to the Sun, the latter cannot be treated as a point source, so we model the number density of the incident photons at a distance $r$ to the Sun as~\cite{EGRET}
\begin{equation}
n_{\gamma}\left(E_{\gamma}, r\right)=\frac{1}{2} n_{\mathrm{B B}}\left(E_{\gamma}\right)\left[1-\sqrt{1-\left(R_{\odot} / r\right)^{2}}\right],
\label{solarphotonfield}
\end{equation}
where $R_{\odot}$ is the radius of the Sun, $n_{\rm BB}(E_{\gamma})=(8\pi/(hc)^{3})E^{2}_{\gamma}/(\exp(E_{\gamma}/k_{\rm B}T)-1)$ is the black-body photon number density per unit energy. One can see that, for $r\gg R_{\odot}$, Eq.~(\ref{solarphotonfield}) reduces to the inverse-square law,
\begin{equation}
n_{\gamma}\left(E_{\gamma}, r\right)\simeq \frac{1}{4} n_{\mathrm{B B}}\left(\frac{R_{\odot}}{r}\right)^2.
\end{equation}

DAMPE's measurement of the CRE spectrum can be fit a broken power law with an excess at $E\sim 1.4\,{\rm TeV}$~\cite{DAMPE_Nature}; we use this as the input to calculate the corresponding solar IC spectrum. The local CRE spectrum (without the excess) is measured to have a double power-law form as
\begin{equation} 
\bar{\Phi}=\Phi_{\mathrm{c}} E^{-\alpha}\left[1+\left(\frac{E_{1}}{E}\right)^{\delta}\right]^{\Delta \alpha_{1} / \delta}\left[1+\left(\frac{E}{E_ {2}}\right)^{\delta}\right]^{\Delta \alpha_{2} / \delta},\label{cre}
\end{equation}
where $\Phi_{\mathrm{c}}=247.2 \,\mathrm{GeV}^{-1} \mathrm{~m}^{-2} \mathrm{~s}^{-1} \mathrm{sr}^{-1}$, $\alpha=3.092$, $\Delta \alpha_{1}=0.096$, $\Delta \alpha_{2}=-0.968$, $\delta=10$, $E_{1}=50 \,\mathrm{GeV}$ and $E_{2}=885.4 \,\mathrm{GeV}$ are the fitted parameter values~\cite{Fan_2018}. 
Here we assume
CREs 
are isotropic
within the heliosphere, and for simplicity, we use a single-bin excess at $\sim 1.4\,{\rm TeV}$ in the following computation.

CREs coming into the heliosphere are subject to the combined effect of outwards solar winds and the surrounding magnetic field, leading to variations in their energy and intensity, known as solar modulation. According to the force field approximation used to obtain the modulated differential CRE intensity~\cite{forcefield}, solar modulation can be described by a one-dimensional potential $\Phi(r)$ relating the CRE spectrum at Earth to any location in the heliosphere:
\begin{equation}
J\left(r, E_{\rm e}\right)=\frac{J\left(\infty, E_{\rm e}+e\Phi(r)\right) \times E_{\rm e}\left(E_{\rm e}+2 E_{0}\right)}{\left(E_{\rm e}+e\Phi(r)+2 E_{0}\right)\left(E_{\rm e}+e\Phi(r)\right)},
\label{J}
\end{equation}
where $J(r,E_{\rm e})$ is the modulated differential CRE intensity, $J(\infty,E_{\rm e}+e\Phi (r))$ is the local interstellar (i.e., unmodulated) CRE spectrum, $E_0=m_{\rm e}c^{2}$ is the rest mass of the electron, and $E_{\rm e}=(\beta-1)E_{0}$ is the kinetic energy of CREs [$\beta=(1-v^{2}/c^{2})^{-1/2}$ being the electron Lorentz factor]. $\Phi (r)$ is the modulation potential, which can be modeled to have time, charge, and rigidity dependence (see e.g. Ref.~\cite{sm_potential} as an example). Because the modulation effect at the TeV-energy range is not large beyond $1^{\circ}$ region from the Sun~\cite{Petrosian_2023}, here we use a spherically symmetric modulation potential for simplicity~\cite{IC_Moskal} 
\begin{eqnarray}
\Phi(r)=\frac{\Phi_{0}}{1.88} \begin{cases}\vspace{1ex} r^{-0.4}-r_{\rm b}^{-0.4}, & r \geq r_{0}, \\  0.24+8\left(r^{-0.1}-r_{0}^{-0.1}\right), & r<r_{0},\end{cases}
\label{eq:Phi1}
\end{eqnarray}
with $\Phi_0=10^{3}\,$MV being the modulation potential at $1~$AU from the Sun, $r_0=10$, and $r_{\rm b}=100$ (in units of AU). With the local CRE spectrum and the modulation potential at the Earth $\Phi_0$, the local interstellar unmodulated CRE spectrum can be recovered. By combining Eqs.~(\ref{J}) and (\ref{eq:Phi1}), one can derive the modulated CRE spectrum at any position in the heliosphere.

\begin{figure}
	\includegraphics[width=8cm]{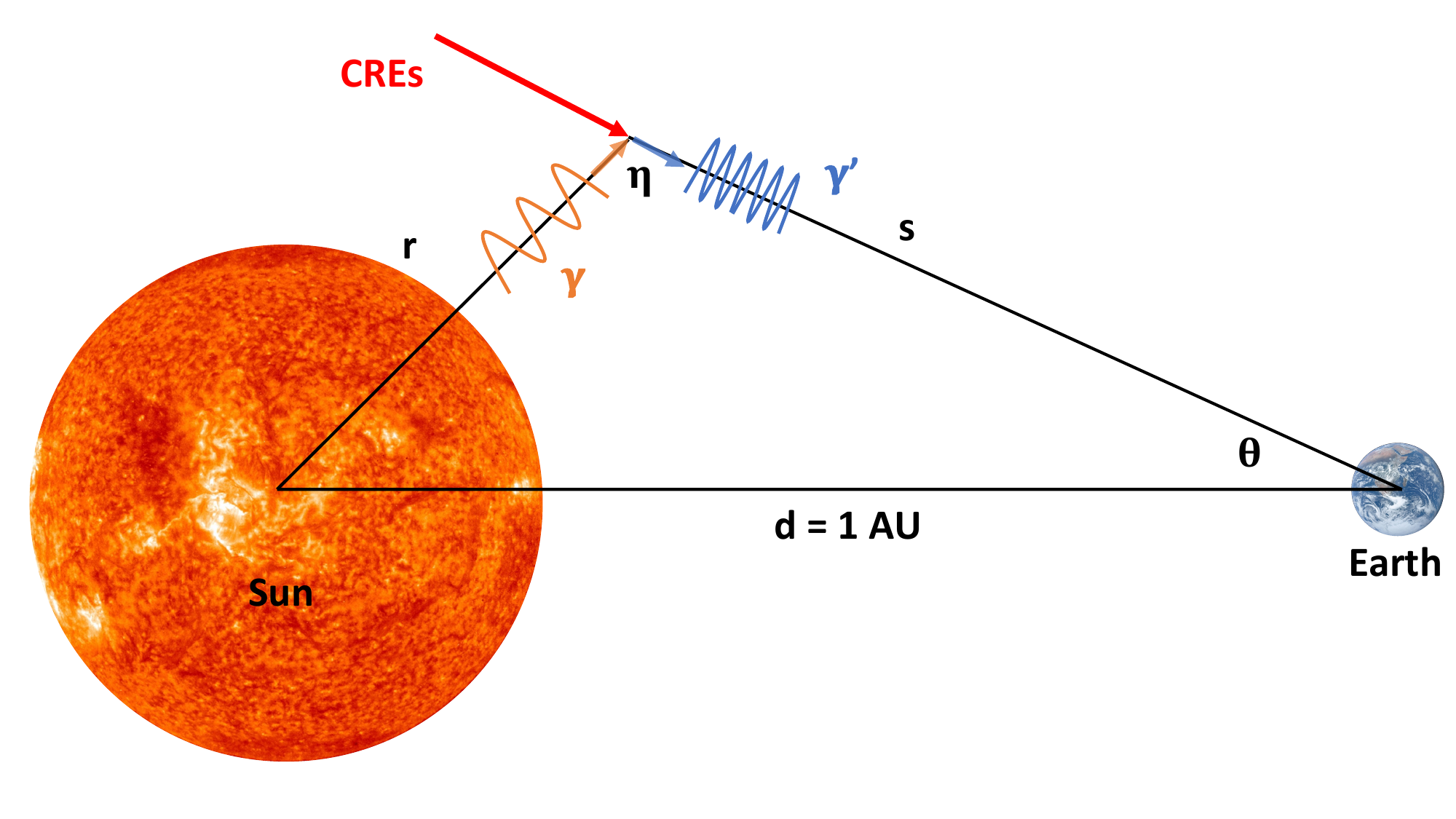}
    \caption{Geometry between CREs, solar radiation field and the observer. Given the very high Lorentz factor ($\beta \gtrsim 10^6$) the IC gamma ray 
    is well-approximated as colinear with the incident CRE.}
    \label{fig:shiyitu}
\end{figure}

The IC emissivity (in units of ${\rm MeV}^{-1}\,{\rm cm}^{-3}\,{\rm s}^{-1}$) at a specific location within the heliosphere can be calculated as
\begin{eqnarray}
\epsilon\left(E'_{\gamma}\right)&&=c\int {\rm d} E_{\rm e} {\rm d}E_{\gamma}\\
&&\times\sigma_{\rm KN}\left(E_{\rm e}, E_{\gamma}, E'_{\gamma}, \eta\right) n_{\gamma}\left(E_{\gamma}, r\right) N\left(E_{\rm e}, r\right), \nonumber\label{eq:epsilon_E}
\end{eqnarray}
where $E_{\rm e}$ is the electron energy, $E_{\gamma}$ the energy of the target photon, $E'_{\gamma}$ the resultant gamma ray photon energy,  $N$ 
and $n_{\gamma}$
are the CRE and target photon number densities per unit energy at the specific location, respectively, $\eta$ is the scattering angle as determined by the geometry relation shown in Fig.~\ref{fig:shiyitu}, and $\sigma_{\mathrm{KN}}$ is the anisotropic Klein-Nishina cross section as given in Ref.~\cite{EGRET},
\begin{eqnarray}
&&\sigma_{\rm KN}\left(E_{\rm e}, E_{\gamma}, E'_{\gamma},\eta\right) = \left(\frac{\pi r_{\rm e}^2 E_0^2}{E_{\gamma} E_{\rm e}^2}\right) \times \left[\left(\frac{E_0}{E_{\gamma}^{\prime\prime}}\right)^2 \left(\frac{\nu}{1-\nu}\right)^2\right. \nonumber\\
&&\left.-  2 \frac{E_0}{E_{\gamma}^{\prime\prime}} \frac{\nu}{(1-\nu)}+(1-\nu)+\frac{1}{1-\nu}\right] \label{eq:sigma_KN} 
\end{eqnarray}
where $E_{\gamma}^{\prime\prime}=\beta E_{\gamma} (1+\cos \eta)$ is the target photon energy in the {\it electron's} rest frame, $r_{\rm e}=e^{2}/m_{\rm e}c^{2}$ is the classical electron radius, and $\nu=E'_{\gamma}/E_{\rm e}$ is the energy transfer fraction. The IC intensity per steradian (in units of ${\rm MeV}^{-1}\,{\rm cm}^{-2}\,{\rm s}^{-1}\,{\rm sr}^{-1}$) is then
\begin{eqnarray}
I(E_{\gamma},\theta)=\frac{1}{4\pi}\int \epsilon (E_{\gamma},s,\theta) \, {\rm d}s, \label{eq:I_def}
\end{eqnarray}
where ${\rm d}s$ is integrated along the line-of-sight, and $\theta$ is the angle from the Sun's center.

\section{Results}
\label{sec:result}

\subsection{Excess feature}
\begin{figure}
	\includegraphics[width=9cm]{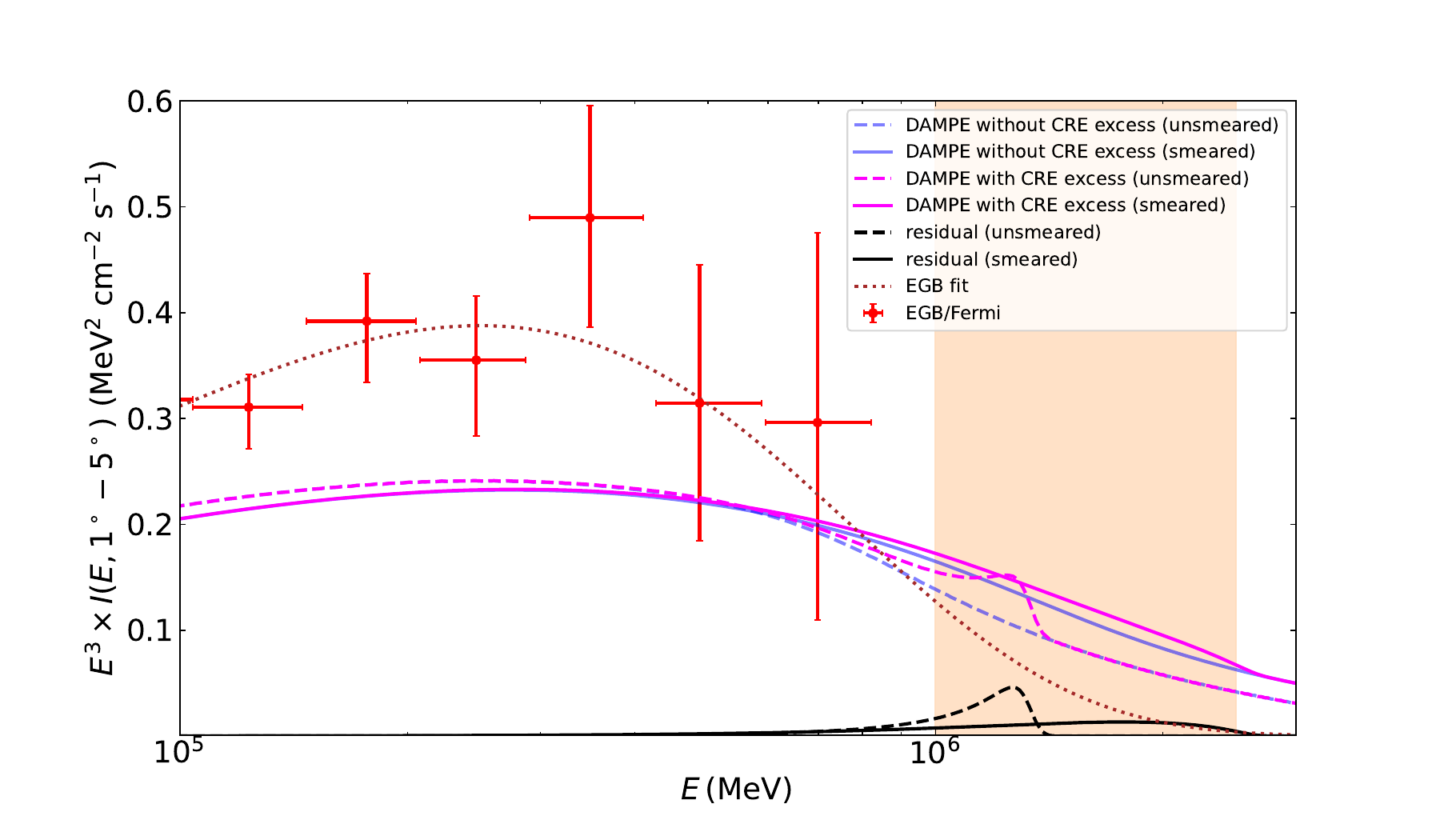}
    \caption{
    Solar IC spectra integrated over the angular area $(1^{\circ},5^{\circ})$. The magenta (blue) dashed line is the Solar IC spectrum given DAMPE spectrum with (without) the $1.4$~TeV excess. The black dashed line represents the net enhancement (difference between magenta and blue). Solid lines show the same result but smeared with a Gaussian function with $\sigma_{E}/E=1$. The orange shaded region indicates the energy integration range used to calculate the significance of the excess detection in Sec.~\ref{sec:result}. The red dots represent the EGB measured by Fermi~\cite{EGB} (integrated over the ring area $1^{\circ}-5^{\circ}$ of the Sun), where the brown dotted line is a power-law fit to the data (Model A in Ref.~\cite{EGB}).}
    \label{fig:result}
\end{figure}

\begin{figure}
	\includegraphics[width=9.5cm]{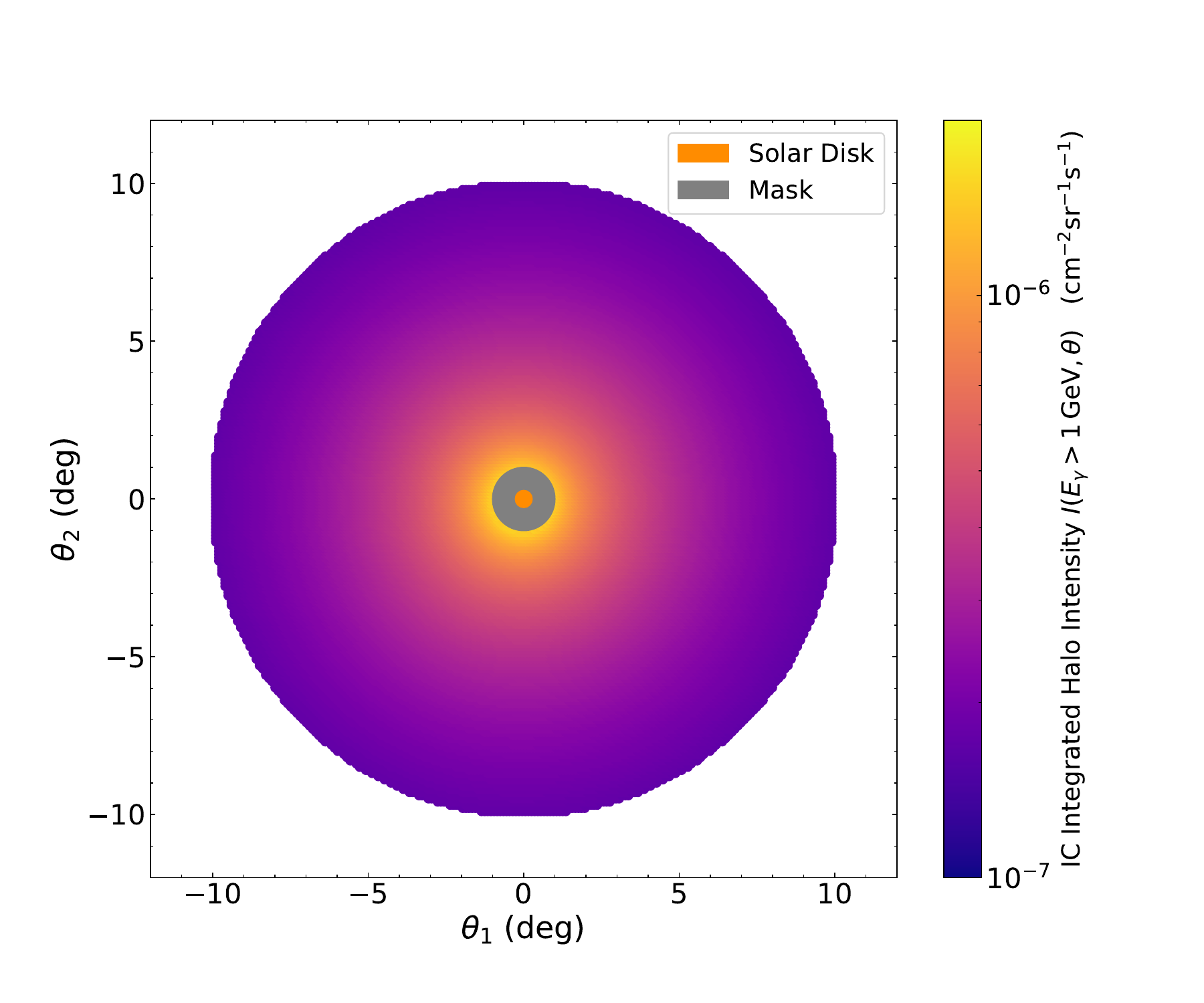}
    \caption{The {\it integrated} halo intensity ($>1\,\mathrm{GeV}$) around the Sun, with $\theta_1$ and $\theta_2$ being the helioprojective longitude and latitude respectively. The orange region represents the angular size of the solar disk. Gamma rays within $1^{\circ}$ region of the Sun are masked out because of the strong contamination brought by the disk component.}
    \label{1117}
\end{figure}

We determine the impact of the 1.4 TeV excess in the
CRE spectrum claimed by DAMPE
on the Solar inverse-Compton (IC) spectrum by employing two different functional representations of the CRE spectrum in our ICS calculation, with and without the excess. 
We use the \texttt{StellarICS} package~\cite{StellarICS} to perform the aforementioned computations. The results of solar IC spectra integrated over the angular area of $1^{\circ}-5^{\circ}$ of the Sun are presented in Fig.~\ref{fig:result} as magenta and blue dashed lines. The black dashed line is the difference between the above two lines, showing the net enhancement around $1\,{\rm TeV}$. In addition, Fig.~\ref{1117} demonstrates the integrated halo intensity of gamma rays with energy $>1 \ \mathrm{GeV}$, depicted as a color map, along with an orange circle indicating the size of the solar disk and a $1^{\circ}$ mask to exclude gamma rays originating from the solar disk direction.

To evaluate the detectability of the excess, it is crucial to account for potential contamination from various sources of gamma ray emission, including Galactic diffuse gamma ray emission (DGE), the extragalactic gamma ray background (EGB), gamma rays originating from the solar disk through hadronic interaction, and point sources. DGE and Milky Way sources can be safely neglected if observing at $|b|>10^{\circ}$. The EGB is the main contamination source for the halo component, especially at a large angular distance from the Sun. The EGB spectrum has been measured by the Fermi telescope from 100 MeV to 820 GeV~\cite{EGB} as shown as the red dots in Fig.~\ref{fig:result}. This can be well modeled by a power law with an exponential cutoff (brown dotted line in Fig.~\ref{fig:result}, see Model A in Ref.~\cite{EGB}),
\begin{eqnarray}
I_{\mathrm{EGB}}=C_0\left(\frac{E}{0.1 \mathrm{GeV}}\right)^{-\mu} \exp \left(-\frac{E}{E_{\mathrm{cut}}}\right), \label{eq:EGB_spectrum}
\end{eqnarray}
where $C_0=1.48 \times 10^{-7} \,\mathrm{MeV}^{-1} \mathrm{~cm}^{-2} \mathrm{~s}^{-1} \mathrm{sr}^{-1}$, $\mu = 2.31$, $E_{\text {cut }}=362 \,\mathrm{GeV}$.

High-energy cosmic ray protons can interact with protons 
of the solar atmosphere, resulting in the production of neutral pions. These pions  decay promptly, leading to the emission of gamma rays from the direction of the solar disk (henceforth the ``disk component'') \citep{SSG1991}. 
%
HAWC detected the 
disk component at $0.5-2.6$ TeV, revealing that the flux of the disk component is approximately one order of magnitude larger than the expected flux of the halo component integrated over an angular area of $\leq 5^\circ$\cite{TeVSun}. Therefore, to enhance the significance of the excess on the solar IC spectrum, 
we apply a $1^\circ$ mask
to exclude the disk component.

Besides future space-borne programs, current ground-based water Cherenkov telescopes, such as HAWC~\cite{HAWC_1} and LHAASO~\cite{ALHAASO} and also the next-generation water Cherenkov telescope SWGO~\cite{SWGO,SWGO2,SWGO3}, possess the capability of detecting TeV gamma rays from the solar direction.
To quantify the detectability of 
an excess
feature in the solar IC spectrum, a large smearing due to the relatively poor energy resolution in the TeV range must be included. We use a Gaussian smearing function with $\sigma_{E}/E=1$ (which is roughly the energy resolution of LHAASO at 1 TeV~\cite{ALHAASO}) to convolve with the predicted signal and background spectra
\begin{equation}
\mathcal{G}(E^{\prime},E)
\begin{cases}
\propto \left(\sqrt{2\pi}\sigma_{E}\right)^{-1}\exp\left(-\frac{(E^{\prime}-E)^2}{2\sigma_{E}^2}\right) &E^{\prime}\leq2E,\\
= 0 &E^{\prime}>2E,
\end{cases}
\end{equation}
where $E^{\prime}$ is the energy of the gamma ray after smearing. The smeared solar IC spectrum is shown in Fig.~\ref{fig:result} by solid lines. By comparing magenta solid and dashed lines (and also black), one can clearly see the ``smoothing out'' feature of the smearing effect, rendering it more challenging to detect the excess.

\subsection{Exposure required}
Water Cherenkov telescopes feature large fields of view and effective areas. The effective area varies with the energy and direction of the incident gamma rays. Therefore, if observing in the direction of the Sun at TeV, its total exposure in a year is determined by the performance of the telescope and its latitude.
The declination of the Sun $\delta_{\odot}$ can be approximated by
\begin{equation}
\delta_{\odot}=-23.44^{\circ} \cdot \cos \left[\left(\frac{360^{\circ}}{365}\right) \cdot(n_{\mathrm{th}}+10)\right]
\label{ds}
\end{equation}
for the $n\mathrm{th}$ day of the year. The zenith angle of the Sun $z_{\odot}$ at a given time satisfies,
\begin{equation}
\cos z_{\odot}=\sin \delta_{\mathrm{tel}} \sin \delta_{\odot}+\cos \delta_{\mathrm{tel}} \cos \delta_{\odot} \cos h,\label{zs}
\end{equation}
where $h$ is the hour angle of the Sun and $\delta_{\mathrm{tel}}$ is the telescope latitude.  LHAASO's total annual exposure 
is
\begin{eqnarray} 
\mathcal{T}_{\mathrm{LHAASO}} &\simeq& \left(A_{15} \times t_{15}+ A_{30} \times t_{30}+ A_{45} \times t_{45}\right. \nonumber\\
&&+ \left.A_{60} \times t_{60}\right)\nonumber\\
&\simeq&  13,894\,\,\mathrm{m}^2\,\mathrm{yr}, 
\end{eqnarray}
where $A_{15}$ represents LHAASO's effective area at 1 TeV within the zenith angle of $15^{\circ}$~\cite{ALHAASO}, $t_{15}$ represents the number of hours in a year that the Sun is within $15^{\circ}$ of the zenith from LHAASO 
using Eqs.~(\ref{ds}) and (\ref{zs}), and similarly for $A_{30}$, $t_{30}$ and the other quantities. Repeating this calculation for HAWC we find $\mathcal{T}_{\mathrm{HAWC}}\simeq 3867\,\,\mathrm{m}^2\,\mathrm{yr} $ (see Ref.~\cite{HAWC} for its effective area for different energies and zenith angles). Notice that, $\mathcal{T}$ in practice could be lower than the derived estimation because of the masking out of the Galactic plane or other point sources that exhibit strong gamma ray emission.

\begin{figure}
    \centering
	\includegraphics[width=9cm]{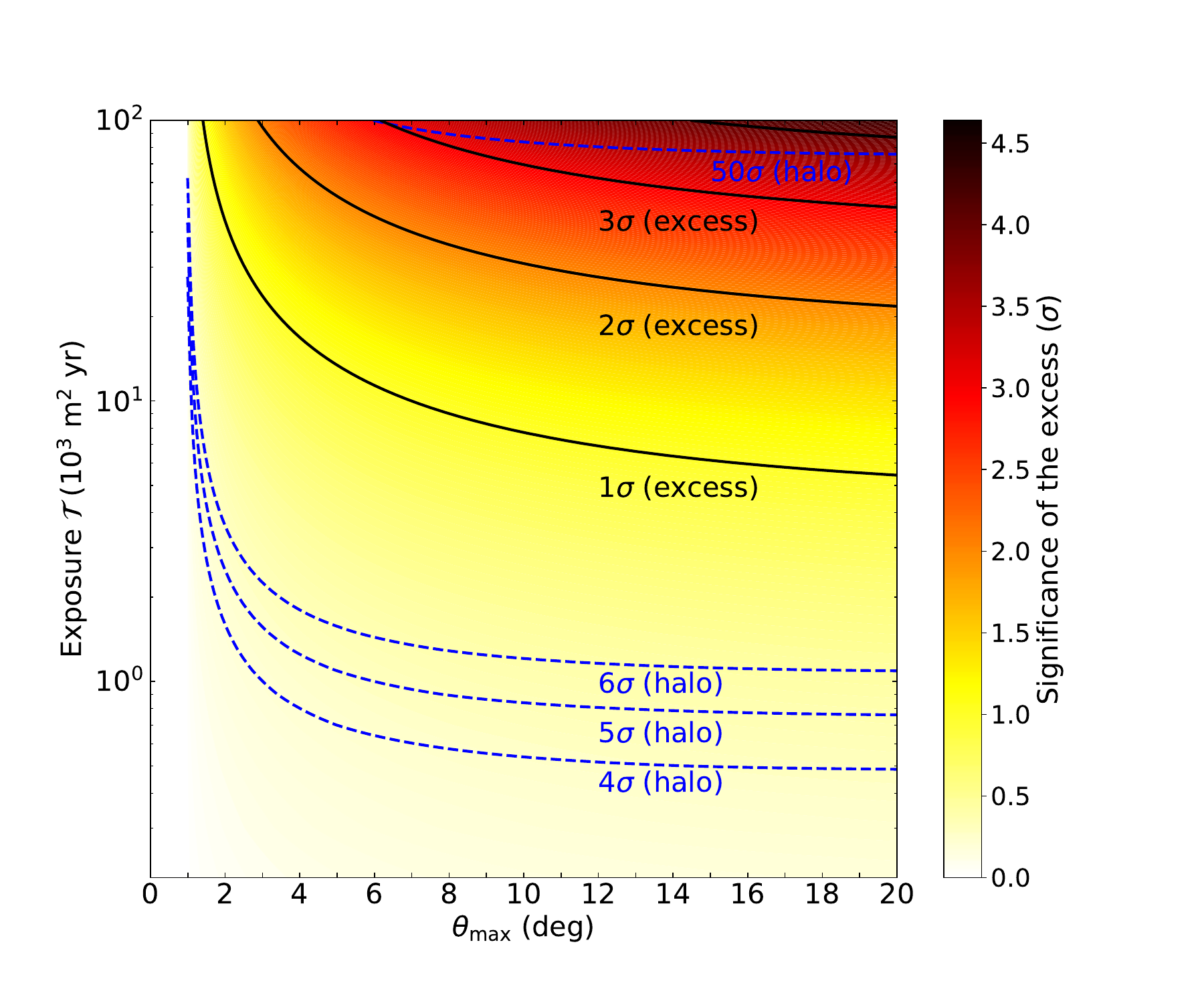}
    \caption{The predicted significance of measurements on the excess of solar IC spectrum at $1-1.4$\,TeV (black lines and color region) and the halo component ($0.5$-$2.5$\,TeV; blue dashed lines). The top black line show the $4\sigma$ detection of the excess signal. With an exposure $\mathcal{T}\simeq 10^{3}\,{\rm m}^{2}\,{\rm yr}$, the halo signal can be measured around $6\sigma$, but cannot achieve excess detection. A larger than $4\sigma$ detection of the excess can only be achieved at $\mathcal{T}\simeq 10^{5}\,{\rm m}^{2}\,{\rm yr}$, while the halo component can be measured at $50\,\sigma$ at this exposure.}
    \label{ob-sig}
\end{figure}

Given a gamma ray telescope observing in an angular range $[\theta_{\rm min}, \theta_{\mathrm{max}}]$ around the Sun, we can calculate the significance 
of an excess measured in the spectrum of the solar halo  IC emission as a function of the exposure time $\mathcal{T}$,
\begin{eqnarray}
S(<\theta_{\mathrm{max}},\mathcal{T}) 
&=& \frac{N_{\rm signal}}{\sqrt{N_{\mathrm{total}}}} \nonumber \\
&=& \int {\rm d}\Omega\int _{E_1}^{E_2} {\rm d}E~I_{\rm signal}(\theta, E) \nonumber \\ 
& \times &\left(\frac{I_{\mathrm{spike}}(\theta,E)+I_{\mathrm{EGB}}(E)}{\mathcal{T}} \right)^{-1/2}, 
 \label{sig}
\end{eqnarray}
where $(E_1, E_2)$ is the energy bin of interest and the solid angle integration is performed within $\theta_{\rm min}<\theta<\theta_{\rm max}$. $I_{\rm signal}$ is the smeared intensity signal we are after, $I_{\rm EGB}$ is the EGB spectrum [Eq.~(\ref{eq:EGB_spectrum})], and $I_{\mathrm{spike}}$ is the smeared intensity of the halo component computed with the single-bin CRE excess (i.e., the magenta solid line in Fig.~\ref{fig:result} divided by $E^3$). The angular integration starts from $\theta_{\rm min}=1^{\circ}$ to safely exclude gamma rays from the disk 
(LHAASO has a resolution of about $0.45^{\circ}$ at 1 TeV~\cite{LHAASO_resolution} and HAWC is about $1^{\circ}$ at 1 TeV~\cite{TeVSun}). We can see that the halo signal can be measured to a relatively high significance if $\mathcal{T}\simeq 10^5\,\mathrm{m}^2\,\mathrm{yr}$. To see this, we substitute $I_{\rm flat}$, which is the smeared intensity of the halo component computed without the single-bin CRE excess (i.e., the blue solid line in Fig.~\ref{fig:result} divided by $E^3$), into $I_{\rm signal}$ and use the energy band $[E_{1}, E_{2}]=[0.5, 2.5]\,{\rm TeV}$ for integral limits in Eq.~(\ref{sig}). We plot the results in Fig.~\ref{ob-sig} as the blue dashed lines. One can see that, with exposure time $\simeq 10^{3}\,{\rm m}^{2}\,{\rm yr}$, the halo component can be measured at around $5$-$6\sigma$ C.L., depending on the maximum angle of observation ($\theta_{\rm max}$). If exposure reaches $\mathcal{T}\simeq 10^{5}\,{\rm m}^{2}\,{\rm yr}$, the significance can reach $50\sigma$ C.L. 
At a given significance level, increasing the integration angular range ($\theta_{\rm max}$) can reduce the demanded exposure time. If we restrict the energy band of interest to  $[E,1.4E]$, i.e., in a narrow band of $\Delta_{E}=0.4 E$ centered at $1.2 E$, and integrate over the solid angle range $\theta\,\in[1^{\circ},10^{\circ}]$, we can obtain an approximate numerical relation for the significance of the halo measurement as a function of exposure time,
\begin{equation}
    S_{\mathrm{halo}}(<10^{\circ},E,\mathcal{T})=\left(\frac{E}{6.6\,\mathrm{TeV}}\right)^{-1.194}\sqrt{\frac{\mathcal{T}}{10^4\,\mathrm{m}^2\,\mathrm{yr}}}.
\end{equation}

We now consider the detection of the excess signal. We substitute $I_{\rm signal}=I_{\rm spike}-I_{\rm flat}$ and $[E_{1}, E_{2}]=[1.0, 2.5]\,{\rm TeV}$ into Eq.~(\ref{sig}) and calculate the significance of the detection as a function of $\mathcal{T}$ and $\theta_{\rm max}$. This energy range safely covers the expected excess signal (the orange shaded region in the lower panel of Fig.~\ref{fig:result}). We show the forecasted significance in Fig~\ref{ob-sig}. We can see that the significance of an 
excess measurement is much smaller than for the halo component at a given exposure. But with $\mathcal{T}=10^5\,\mathrm{m}^2\,\mathrm{yr}$, the DAMPE-motivated excess in the solar IC spectrum can be detected with better 
than $4\sigma$ C.L. Given the yearly exposures of LHAASO and HAWC, we 
estimate that this level of detection is achievable by LHAASO in $10^{5}/13,894\,{\rm yr}\simeq  7.2\,{\rm yrs}$, and HAWC in $10^{5}/3,867\,{\rm yr}\simeq  25.9\,{\rm yrs}$. Therefore, we conclude that it is feasible to use the solar halo IC spectrum to cross-check TeV features in the local CRE spectrum with long exposures of water Cherenkov telescopes like HAWC~\cite{HAWC_1} and LHAASO~\cite{ALHAASO}. 
We anticipate, in addition, that such observations can provide constraints
on possible spectral features in the CRE  distribution at the energy scales
beyond current practicable, direct measurements.

\section{Conclusion}
\label{sec:conclude}
An independent check exploiting the solar inverse Compton halo emission
can help test the robustness of any claimed
feature detected via direct measurements of the cosmic ray electron and positron spectrum. 
In this paper, we take the CRE excess at $1.4\,{\rm TeV}$ measured by DAMPE as an example to calculate the predicted off-disk solar emission due to  inverse Compton-scattering. We derive the IC spectrum with and without the $1.4\,{\rm TeV}$ excess in the CRE spectrum, and show an expected enhancement of solar IC intensity at $\sim 1$-$1.4\,\mathrm{TeV}$. We then forecast the detectability of this excess signal, and the halo component itself, by including the contamination brought by the extragalactic gamma ray background. We show that with $10^{3}\,{\rm m}^{2}\,{\rm yr}$ total exposure, the halo component can be measured at $5$-$6\sigma$ C.L., but to detect the excess signal in the solar IC spectrum at $\gtrsim 4\sigma$ C.L. the total exposure is required to reach $\mathcal{T}=10^5\,\mathrm{m}^2\,\mathrm{yr}$ in the off-disk direction. 
Using
the effective areas for HAWC~\cite{HAWC_1} and LHAASO~\cite{ALHAASO} (water Cherenkov telescopes), we show that the excess signal can be  detected with $25.9\,{\rm yrs}$ observations of HAWC and $7.2\,{\rm yrs}$ of LHAASO. Our result shows the feasibility of testing a single-bin excess in the CRE spectrum (as motivated by the DAMPE excess) using the solar IC spectrum.

\begin{acknowledgments}
We would like to thank Chris Gordon for helpful discussions. This work is supported by the National Natural Science Foundation of China (12047503, 12275278), the National Research Foundation of South Africa under Grants No.~150580, No.~120385, and No.~120378, and NITheCS program ``New Insights into Astrophysics and Cosmology with Theoretical Models confronting Observational Data''.
\end{acknowledgments}

\bibliography{apssamp}

\end{document}